\documentclass[paper, 12pt, letterpaper]{JHEP} \input{epsf.tex}
\usepackage{graphics} \usepackage{epsfig}
\usepackage{amsmath}\usepackage{amssymb}

   \def\Z{{\mathbb
    Z}}

\def\id{\protect{{1 \kern-.28em {\rm l}}}}

\newcommand{\be}{\begin{equation}} \newcommand{\ee}{\end{equation}}
\newcommand{\bea}{\begin{eqnarray}} \newcommand{\eea}{\end{eqnarray}}
\newcommand{\beann}{\begin{eqnarray*}}
  \newcommand{\eeann}{\end{eqnarray*}}
\newcommand{\bfig}{\begin{figure}} \newcommand{\efig}{\end{figure}}
\newcommand{\nn}{\nonumber}
\newcommand{\ba}{\begin{array}}\newcommand{\ea}{\end{array}}

\newtheorem{Proposition}{Proposition}[section]

\newtheorem{Theorem}{Theorem}[section]
\newtheorem{Lemma}{Lemma}[section]
\newtheorem{Corrolary}{Corrolary}[section]

\newcommand{\bp}{\begin{Proposition}}
  \newcommand{\ep}{\end{Proposition}}
\newcommand{\bt}{\begin{Theorem}} \newcommand{\et}{\end{Theorem}}
\newcommand{\bl}{\begin{Lemma}} \newcommand{\el}{\end{Lemma}}
\newcommand{\bc}{\begin{Corrolary}} \newcommand{\ec}{\end{Corrolary}}

 \newcommand{\eps}{\varepsilon}

\newcommand{\void}[1]{}

   \def\ep{\eps}

\title{Geometric regularizations and dual conifold transitions}

\author{K. Landsteiner, C. I. Lazaroiu\\ 
Insitut f\"ur Physik\\
Humboldt Universit\"at zu Berlin\\ 
Invalidenstrasse 110, Berlin\\ Germany\\
landsteiner, calin@physik.hu-berlin.de}

\abstract{We consider a geometric regularization for the class of conifold
transitions relating D-brane systems on noncompact Calabi-Yau spaces to
certain flux backgrounds. This regularization 
respects the $SL(2,\Z)$ invariance of the flux
superpotential, and allows for computation of the relevant periods through the
method of Picard-Fuchs equations.  The regularized geometry is a
noncompact Calabi-Yau which can be viewed as a monodromic fibration, with the
nontrivial monodromy being induced by the regulator. It reduces to the
original, non-monodromic background when the regulator is removed. Using this
regularization, we discuss the simple case of the local conifold, and show how
the relevant field-theoretic information can be extracted in this approach.}

\preprint{HU-EP 03/14}

\begin{document}

\tableofcontents

\section{Introduction}

Realizing supersymmetric gauge theories in string theory with the help of
D-branes has lead to progress in understanding their non-perturbative 
dynamics. A recent prominent example is ${\cal N}=1$ $U(N)$ gauge
theory with an additional chiral multiplet in the adjoint representation and a
superpotential of the form \cite{Vafa_ts, Cachazo}:
\be
W(\Phi) = \sum_{k=1}^{n+1} \,\frac{t_k}{k}\, \mathrm{tr} ( \Phi^{k}).
\ee  
Based on the results in \cite{Gopakumar:1998ki} it has been argued that this 
can be geometrically engineered in type IIB string theory with the
help of the non-compact Calabi-Yau space:
\be
\label{cy}
W'(x)^2 + f_{n-1}(x) + y^2 + s^2 + t^2 =0\,,
\ee 
where $W$ is the function defined by the superpotential and $f_{n-1}$ is a
polynomial of degree $n-1$. When turning off these deformations, the space 
(\ref{cy}) acquires
singular points sitting above the the $n$ roots $x_i$ of $W'(x)=0$. 
The gauge theory interpretation 
arises by blowing up these singularities to $\mathbb{P}^1$'s and partially
wrapping $N_i$ D5-branes on them. The low energy dynamics 
on the D-branes describes confining vacua of the ${\cal N}=1$ 
supersymmetric gauge theory. This is reflected in a geometric transition
where the resolved geometry is replaced by the deformed one (\ref{cy}).
The D-branes disappear in this process, being replaced by three-from
flux through the three-cycles of the deformed geometry. 
The gauge theory has vacua in which the vev of
$\Phi$ is $\langle \Phi \rangle = \mathrm{diag}(x_1 {\bf 1}_{N_1},\dots, x_n
{\bf 1}_{N_n})$ and the gauge group is broken to the product
$\prod_{i=1}^n U(N_i)$. At low energies the non-Abelian parts of the broken
gauge group will confine and for each gauge group factor there will be a
gaugino condensate $S_i$. The effective superpotential 
for these condensates can be computed from the Calabi-Yau geometry with the 
help of its periods:
\be
S_i = \int_{A_i} \Omega ~~\,,\,~~ \Pi_i = \frac{\partial F}{\partial S_i} =
\int_{B_i} \Omega ~~,
\ee
where $\Omega$ is the holomorphic three form of the Calabi-Yau space
and $A_i,B_i$ give a canonical basis of 3-cycles. This 
superpotential takes the form \cite{flux1, flux2, flux3, flux4}:
\be
\label{W_flux}
-\frac{1}{2\pi i} W_{\mathrm{eff}} = \sum_{i=1}^n (N_i \Pi_i + \alpha_i S_i)\,,
\ee
where $N_i,\alpha_i$ are the fluxes of the 
type IIB three form $H_R+\tau H_{NS}$ through $A_i$ and $B_i$.  The
gauge theory interpretation identifies $N_i$ with the rank of the
$i^\mathrm{th}$ factor of the unbroken gauge group and $\alpha_i$ with the
bare coupling of this factor group. The $n$ coefficients of the polynomial
$f_{n-1}(x)$ can be determined in terms of the gaugino condensates
$S_i$. Finally, by integrating over $s,t$ one can reduce the
Calabi-Yau space to the affine algebraic curve:
\be
\label{nonregsurface}
y^2 +W'(x)^2 + f_{n-1}(x)=0 \,,
\ee 
while the holomorphic three-form descends to the following meromorphic 
differential on this Riemann surface\footnote{The papers \cite{Cachazo,
    Cachazo2, Cachazo3} use a different convention for $\omega$, which amounts
  to dropping a prefactor of $1/2$ at the price of integrating over only
  half of the length of each Riemann surface cycle. In this paper, we shall 
  always integrate
  along the full cycles on the Riemann surface.} :
\be
\label{omega}
\omega = \frac{i}{2}y\, dx~~.
\ee
Recently this string theoretic setup has lead to the conjecture that the effective superpotential 
of the gauge theory can be computed by using a matrix model whose action is given
by the tree-level superpotential. The Riemann surface (\ref{nonregsurface})
has a matrix model interpretation as a spectral curve
\cite{matrix1, matrix2, matrix3, matrix4, matrix5, matrix6, matrix7, matrix8}.

One of the interesting features of this construction is that in the geometry
(\ref{cy}) only the $A$-cycles are compact. The `$B$-cycles' are non-compact
\footnote{More precisely, only $n-1$ $B$-cycles can be chosen to be compact.
The missing `cycle' is then a curve defined 
in terms of a cutoff regulator. This cutoff construction will be recalled
below for the simple case of a local conifold.}, a
feature which forces one to introduce a cutoff $\Lambda_0$ in the 
$B$-period integrals \cite{Cachazo, Cachazo2}. Related to this is the fact 
that the meromorphic form (\ref{omega}) is a differential of the third kind, 
having  a nonzero residue at $x=\infty$.
This is necessary since otherwise the number of independent
$A$-periods would be only $n-1$, in disagreement with the number of gaugino 
condensates. Also notice that this regularization breaks the $SL(2,\Z)$ 
invariance of the flux superpotential (\ref{W_flux}). 

For the gauge theory interpretation the cutoff regularization is not
unwellcome, since it is used to renormalize the bare gauge couplings
$\alpha_i$. However, it is interesting to ask what happens if one picks 
a geometric regularization instead.  In particular, 
it could prove convenient for some applications to
use a regularization which preserves the $SL(2,\Z)$ invariance of
the flux superpotential (\ref{W_flux}).

This is the question we wish to study in the present note.  The
geometric regularization we shall choose will compactify the $B$-cycles while
promoting (\ref{omega}) to a meromorphic differential of the second kind on
the modified algebraic curve.  
The result will be a closed Riemann surface, whose periods can be computed
in standard manner with the help of Picard-Fuchs equations.

The simplest regularization satisfying
our  requirements is a small deformation of the fibered geometry (\ref{cy}) 
which
transforms it into a monodromic fibration in the sense of \cite{Cachazo3}.
This fibration will have supplementary conifold degenerations 
when compared with the original geometry. 
While the resolutions of these singular limits need 
not admit a simple interpretation in terms of 
partially wrapped $D5$-branes, the modified geometry does 
make perfect sense as a string 
background. On the deformation side, this is still a 
non-compact background with fluxes, thereby leading to the well-known flux 
superpotential of \cite{flux1, flux2, flux3, flux4}. One can take the
limit $\alpha'\rightarrow 0$ while keeping the geometric regulator 
fixed. This leads to an effective 
four-dimensional description which inherits the 
$SL(2,\Z)$ 
duality of the the original type IIB string, a property which is reflected in
the manifestly $SL(2,\Z)$ invariant form of the flux superpotential.
As in \cite{Vafa_ts}, one can decouple gravity by focusing on {\em one}
Calabi-Yau singularization and identifying its vanishing periods with the
relevant gaugino condensates. Assuming that there exists 
an $SL(2,\Z)$ transformation
which makes all vanishing cycles carry RR flux, one then takes the limit
in which these fluxes $N_i$ become large, while the string coupling
computed in that $SL(2,\Z)$ frame vanishes such that the quantities
$g_s N_i$ stay fixed.

In this note, we shall focus on the simplest
situation, namely when the gauge group remains unbroken, the superpotential is
quadratic in $\Phi$ and the algebraic curve (\ref{nonregsurface}) has a single
cut in the $x$ plane. Modifying the geometry will lead to a 
smooth complex torus. We compute the periods by solving the
associated Picard-Fuchs equation and give a discussion of the physics that
emerges when instead of the $A$ cycle, it is the $B$ cycle which becomes
small. We point out some possible generalizations in the last section.

\section{The cutoff regularization}

\begin{figure}[hbtp]
\begin{center}
\scalebox{0.5}{\input{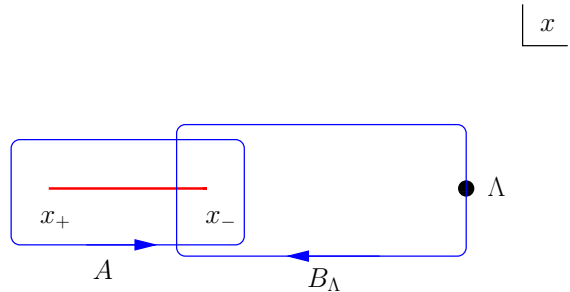}}
\end{center}
\caption{Branch-cut for the undeformed curve. }
\label{cuts}
\end{figure}

The non-compact Calabi-Yau threefold corresponding to the simplest gauge
theory vacuum with unbroken $U(N)$ gauge group is the local conifold:
\be
\label{conifold}
x^2 + y^2 + s^2 + t^2 + \mu =0\,.
\ee 
After integrating over the $(s,t)$ coordinates, this leads to the algebraic
curve:
\be
\label{curve}
y^2+x^2+\mu=0~~.
\ee
Projectivizing (\ref{curve}) gives a hyperelliptic curve $C$ in $\mathbb{P}^2$,
described by the equation:
\be
\label{proj_curve}
Y^2+X^2+\mu Z^2=0~~,
\ee
where $X,Y,Z$ are the homogeneous coordinates.  The projective curve
(\ref{proj_curve}) develops an ordinary double point at the origin for
$\mu=0$.  For $\mu\neq 0$, this curve is a smooth Riemann surface of genus
{\em zero}, i.e. a copy of $\mathbb{P}^1$.  The quantity $y$ has a single cut which
connects the points $x_\pm=\pm i \sqrt{\mu}$ (figure \ref{cuts}).

\begin{figure}[hbtp]
\begin{center}
\scalebox{0.5}{\input{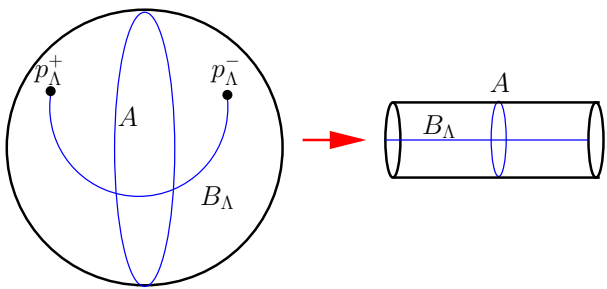}}
\end{center}
\caption{Geometric interpretation of the cutoff regularization. }

\label{cyl}
\end{figure}

The projectivized curve has two points above $x=\infty$, which are obtained by
setting $Z=0$ in its defining equation.  These are the points
$p^\pm_\infty=[1,\pm 1,0]\in \mathbb{P}^2$, each sitting on one of the branches of
(\ref{proj_curve}).  The cutoff regularization of \cite{Cachazo} 
replaces these with two points sitting at a finite distance along the 
$x$-plane.  Let us give a precise description of this regularization.
Picking a complex number $\Lambda_0$ (with $|\Lambda_0|>>1$), the curve
(\ref{proj_curve}) has two points sitting above $x=\Lambda_0$, namely 
$p^\pm_{\Lambda_0}= [\Lambda_0, \pm i\sqrt{\Lambda_0^2+\mu^2},1]$.  
Removing these
from the projectivized curve gives a twice punctured sphere ${\tilde
C}=C-\{p^+_{\Lambda_0},p^-_{\Lambda_0}\}$, which is conformally equivalent  
with an infinite cylinder
(figure \ref{cyl}).  The geometric regularization of \cite{Cachazo} amounts
to working with this twice-punctured sphere instead of the curve
(\ref{proj_curve}). The generator of $\pi_1({\tilde C})=\Z$ plays the role of
$A$-cycle, while the `B-cycle' $B_{\Lambda_0}$ of \cite{Cachazo} is an open 
{\em path} connecting the points $p^\pm(\Lambda_0)$ sitting in the conformal 
compactification $C$ of ${\tilde
C}$. Hence the regularized B-period $\int_{B_{\Lambda_0}}\omega$ of
\cite{Cachazo} is a sort of `holomorphic length' of the cylinder
${\tilde C}$.

In projective coordinates, the differential $\omega$ takes the form:
\be
\label{lambda_proj}
\omega = \frac{i}{2}\frac{Y}{Z}\, d \left(\frac{X}{Z}\right) = 
\frac{i}{2}\left(\frac{1}{Z^2} Y dX -\frac{1}{Z^3} XYdZ\right)~~.
\ee
To study the behavior at infinity we can go to the coordinate patch 
$X=1$ where the curve takes the form
$ Y^2 +\mu Z^2+1=0$. For small $Z$, we find:
\be
\omega = -\frac{dZ}{2Z^3}\sqrt{1+\mu Z^2} = -\frac{dZ}{2Z^3} - 
\frac{\mu}{4} \frac{dZ}{Z} + O(Z)\, dZ \,,
\ee
which makes the pole with residue $-\mu/4$ explicit. 
The A-type period can in this case
be simply computed as the negative of the residue of $\omega$ at $Z=0$.

\

\section{Geometric  regularization of the deformed conifold}

We shall replace (\ref{curve}) with the `regularized' curve:
\be
\label{reg_curve}
y^2+\epsilon x^3+x^2+\mu=0~~,
\ee
where we take $\epsilon$ to be a small complex quantity.  Correspondingly, we
replace the local conifold (\ref{conifold}) with the affine Calabi-Yau
threefold:
\be
\label{reg_conifold}
\epsilon x^3 + x^2 + y^2 + s^2 + t^2 +\mu=0~~.
\ee
This can be viewed as a monodromic $A_1$ fibration over the $x$-plane, 
in the sense of \cite{Cachazo3}. As in \cite{Klemm:1996bj, Cachazo, Cachazo2, Cachazo3}, 
one can integrate the holomorphic 3-form:
\be
\Omega=\frac{i}{2\pi}\frac{dx\wedge dy \wedge ds}{t}=
-\frac{i}{2\pi}\frac{dx\wedge dy \wedge dt}{s}= 
\frac{i}{2\pi}\frac{dx\wedge ds \wedge dt}{y}=-\frac{i}{2\pi}
\frac{2dy \wedge ds \wedge dt}{3\epsilon x^2+2x}~~
\ee
over the fiber coordinates $s,t$ in order to reduce it to the 
meromorphic 1-form $\omega=\frac{i}{2}ydz$ 
on the Riemann surface (\ref{reg_curve}). 
This is achieved by choosing 3-cycles which are obtained by fibering certain 
two-spheres sitting inside the $s,t$ fibers over a curve in the $x$ plane. 

\begin{figure}[hbtp]
\begin{center}
\scalebox{0.5}{\input{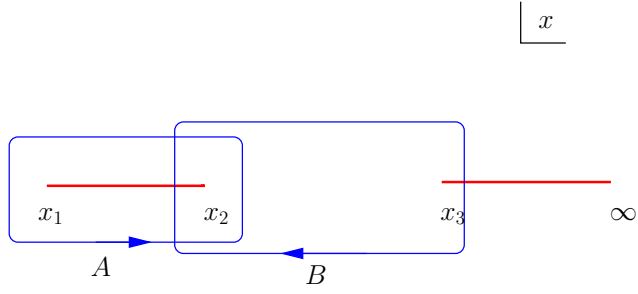}}
\end{center}
\caption{Cuts of the geometrically regularized curve for $|\epsilon^2\mu|<<1$.}
\label{reg_cuts}
\end{figure}

For $|\epsilon^2\mu|<<1$, the $x$-polynomial in (\ref{reg_curve}) has three 
zeroes, namely:
\bea
x_1&=&x_+ +\frac{1}{2}\mu\epsilon +O(\epsilon^2\mu)\nn\\
x_2&=&x_- +\frac{1}{2}\mu\epsilon +O(\epsilon^2\mu)\\ 
x_3&=&-\frac{1}{\epsilon}+O(1)~~.\nn
\eea 
Hence the geometric regularization introduces a new cut connecting $x_3$ and
the point at infinity, while performing a small displacement of the cut of the
original curve (\ref{curve}). In particular, we now have a 
compact $B$-cycle encircling $x_2$ and $x_3$. 
This situation is shown in figure \ref{reg_cuts}.

\begin{figure}[hbtp]
\begin{center}
\scalebox{0.5}{\input{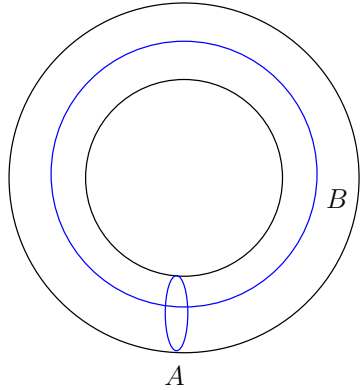}}
\end{center}
\caption{The geometric regularization produces a complex torus.}
\label{torus}
\end{figure}

We next consider the projectivization of (\ref{reg_curve}), which has the
form:
\be
\label{proj_reg_curve}
Y^2 Z+\epsilon X^3+X^2 Z+\mu Z^3=0~~.
\ee
This projective curve has genus $1$, hence it describes a complex torus
(figure \ref{torus}). It develops singularities for $\mu=0$ (when the cycle
$A$ collapses to an ordinary double point, sitting at $[0,0,1]$) and
$\mu=-\frac{4}{27\epsilon^2}$ (when the $B$-cycle is pinched to an ODP sitting
at $[-\frac{2}{3\epsilon},0,1]$).  The curve (\ref{proj_reg_curve}) has a
single point sitting above $x=\infty$, namely $p_\infty=[0,1,0]$.

\subsection{The dual degenerations}

\begin{figure}[hbtp]
\begin{center}
\scalebox{0.5}{\input{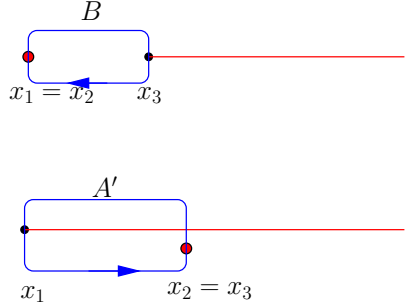}}
\end{center}
\caption{The degenerations $\mu=0$ (above) and $\mu=-\frac{4}{27\epsilon^2}$
(below). In the second figure, we have slightly displaced the point $x_2=x_3$
for clarity.}
\label{cdegs}
\end{figure}

Let us give a more detailed discussion of the degenerations of
(\ref{reg_curve}) for $\mu=0$ and $\mu=-\frac{4}{27\epsilon^2}$. 
Consider the polynomial:
\be
\label{p}
p(x)=\epsilon x^3 + x^2 + \mu~~. 
\ee
For $\mu=0$, this 
factors as $p(x)=\epsilon x^2(x+1/\epsilon)$. Hence the roots $x_1$ and $x_2$ 
coalesce in this limit, which means that the cut $[x_1,x_2]$ reduces to a 
double point (figure \ref{cdegs}). The degenerate Riemann surface has branches:
\be
\label{b1}
y_\pm(x)=\pm i x\sqrt{\epsilon x+1}~~,
\ee
which are interchanged by the monodromy around $x_3$.

For $\mu=-\frac{4}{27\epsilon^2}$, we have 
$p(x)=\epsilon(x-\frac{1}{3\epsilon})(x+\frac{2}{3\epsilon})^2$. In this case,
$x_2$ and $x_3$ have coalesced to a double point, while $x_1$ is connected
to $\infty$ by a branch cut (see figure \ref{cdegs}). The degenerate 
surface has branches:
\be
\label{b2}
y_\pm(x)=\pm i (x+\frac{2}{3\epsilon})\sqrt{\epsilon x-\frac{1}{3}}~~.
\ee
These are interchanged by the monodromy around $x_1$. 

Also notice that the second degenerate curve can be obtained from the first
by performing the transformation:
\bea
\label{isomf}
x&\rightarrow& -x-\frac{2}{3\epsilon}\nn\\
y&\rightarrow& -iy~~,
\eea
which clearly maps (\ref{b1}) into (\ref{b2}). This change of coordinates 
maps the ODP of the first degenerate curve into that of the second
curve, while interchanging the cuts. Thus (\ref{isomf}) identifies the two 
degenerations, while mapping the $B$ cycle of the first into the $A$ 
cycle of the second. The $B$-period 
$\frac{i}{2}
\int_B{\frac{dx}{2\pi i} y}$ of the first degeneration is then mapped to 
$-i$ times the $A$-period $\frac{i}{2}\int_{A'}{\frac{dx}{2\pi i} y}$
of the second degeneration. As we shall see in more detail below, this 
symmetry can be viewed as a remnant of the $SL(2,\Z)$ symmetry of 
the type IIB flux background on the geometrically regularized space 
(\ref{reg_conifold}).

For each conifold singularization of (\ref{reg_conifold}), one can see that 
the two-sphere 
obtained by a small resolution will not be monodromy invariant.
As explained in \cite{Cachazo3}, this prevents us from wrapping $D$-branes 
on such a sphere in the resolved geometry, which means that the low energy 
limit of the type IIB background on the 
deformed space (\ref{reg_conifold}) does not admit a simple gauge theory 
description. To recover a standard gauge-theoretic interpretation, one must 
take the limit $\epsilon\rightarrow 0$. This is 
exactly what one expects based on the arguments of \cite{flux2}. 
Keeping $\mu$ finite and small, 
the limit $\epsilon\rightarrow 0$ has the effect 
of pushing the branch point $x_3$ toward infinity, thereby replacing the cut 
$[x_3, \infty]$ with an ordinary double point at $x=\infty$. In this 
limit, the conifold point at $\mu=-\frac{4}{27\epsilon^2}$ is pushed 
to infinity in the moduli space. The regularized curve (\ref{reg_curve}) 
tends to the the original curve (\ref{curve}) uniformly over compact domains 
in the $x$-plane. However we note that, 
starting with the regularized model, one can take 
a different limit, namely $\epsilon\rightarrow 0$ while 
$\epsilon^2\mu$ is kept fixed and such that 
$|1+\frac{27\epsilon^2\mu}{4}|$ is small. As we shall see below, this 
is equivalent with the previous limit through an $SL(2,\Z)$ transformation.

\

The differential $\omega$ on the regularized curve (\ref{proj_reg_curve})
is given by (\ref{lambda_proj}). In the patch $X=1$ we now have a branch point
at $Z=0$ and for this reason we introduce the coordinate $\zeta^2=Z$ which is
single valued around this point.  Expanding $\omega$ for small $\zeta$ gives:
\be
\omega = -\sqrt{\epsilon}\frac{d\zeta}{\zeta^6} - 
\frac{\mu}{2\sqrt{\epsilon}} \frac{d\zeta}{\zeta^4} +
\frac{\mu^2}{8 \epsilon^{3/2}}\frac{d\zeta}{\zeta^2} + O(\zeta^0)d\zeta   \,,
\ee
showing that $\omega$ is an Abelian differential of the second kind on
(\ref{proj_reg_curve}).

\

\section{Periods of the geometrically regularized surface}

In this section, we shall extract the $(A,B)$ periods of the meromorphic form
$\omega=\frac{i}{2}ydz$ 
for the regularized curve (\ref{reg_curve}). Since the latter is
a closed Riemann surface, one can use standard techniques in order to write
down a Picard-Fuchs equation for the periods, and extract their moduli
dependence by solving this equation.

\subsection{The Picard-Fuchs equation}

Let us introduce the rescaled quantities: \bea
\label{scalings}
x=\frac{z}{\epsilon}~~,~~y=\frac{w}{\epsilon}~~,~~
\mu=-\frac{4}{27\epsilon ^2}\nu~~.\nn \eea 
In terms of these variables, the
defining equation (\ref{reg_curve}) becomes: \be
\label{scaled_curve}
w^2+z^3+z^2-\frac{4}{27}\nu=0~~.  \ee

The discriminant of the polynomial $p(z)=z^3+z^2-\frac{4}{27}\nu$ takes the
form:
\be
\Delta=-\frac{16}{27}\nu(\nu-1)~~.
\ee
For $\nu=0$, the elliptic curve (\ref{scaled_curve}) acquires an ordinary
double point (ODP) at the origin, while for $\nu=1$ it develops an ODP at
$z=-2/3$ and $w=0$. These correspond to the degenerations discussed in the 
previous section. 

Under the redefinitions (\ref{scalings}), the form $\omega=\frac{i}{2}ydx$ 
scales as:
\be
\omega=\frac{1}{\epsilon^2}\kappa~~,
\ee
where $\kappa:=\frac{i}{2}wdz$. 
This is a meromorphic differential of the second kind 
on the complex torus
(\ref{scaled_curve}). Thus its 
periods $U=\int_{{\cal C}}{\kappa}$ (with ${\cal C}$
a cycle on the torus) satisfy a Picard-Fuchs equation, which can be
extracted with the methods of \cite{Albrecht} (see \cite{PF} for 
a systematic approach which can be easily coded):
\be
\label{PF}
\nu(\nu-1)\frac{d^2 U}{d\nu^2}+\frac{5}{36} U=0~~.
\ee 
Introducing the logarithmic derivative $\delta=\nu \frac{d}{d\nu}$, one can
write this in the form:
\be
\left[\delta(\delta-1)-\nu(\delta-1/6)(\delta-5/6)\right] U=0~~,
\ee
which can be recognized as a hypergeometric equation with symbol
${\footnotesize \left[\begin{array}{ccc}-1/6&~&-5/6 \\~&0&~ \end{array}
\right]}$.

\subsection{A period basis}

To extract a basis of periods, we shall use the Meijer function technique
described in \cite{periods,periods2}. As explained in that reference, a basis
of solutions of (\ref{PF}) is provided by the following functions,
which we write in terms of their Mellin-Barnes representations:
\bea
U_1(\nu)&=&\frac{1}{2\pi i \Gamma(-1/6)\Gamma(-5/6)}
\int_{\gamma_1}{\frac{\Gamma(-s)\Gamma(s-1/6)\Gamma(s-5/6)}{\Gamma(s)}(-\nu)
^s}\\
U_2(\nu)&=&\frac{1}{2\pi i \Gamma(-1/6)\Gamma(-5/6)}
\int_{\gamma_2}{\frac{\Gamma(1-s)
\Gamma(-s)\Gamma(s-1/6)\Gamma(s-5/6)}{\Gamma(s)}\nu^s}~~.\nn
\eea
Here $\gamma_j$ are contours connecting $-i\infty$ and $+i\infty$ while
separating $(A)$ and $(B)$-type poles of the corresponding integrands.  For
$U_1$, the $(A)$-poles are $s=n$ from $\Gamma(-s)$, while the $(B)$-poles are
$s=-n+1/6$ and $-n+5/6$ from $\Gamma(s-1/6)$ and $\Gamma(s-5/6)$, where $n$ is
a non-negative integer. For $U_2$, the $(A)$-poles are $s=n$ , while the
$(B)$-poles are $s=1/6-n$ and $s=5/6-n$. All poles are simple except for the
nonzero $(A)$-poles of the $U_2$-integrand, which are double.

For $|\nu|<1$, one can close the contour $\gamma_1$ toward $+\infty$ to find:
\be
\label{U_10}
U_1(\nu)=\frac{5}{36} \nu 
~_2F_1\big({\scriptsize\begin{array}{ccc}1/6&~&5/6 \\~&2&~ \end{array}}\big|\nu \big)=
\nu \frac{d}{d\nu} ~_2F_1\big({\scriptsize
\begin{array}{ccc}-1/6&~&-5/6 \\~&1&~ \end{array}}\big|\nu\big) ~_{\stackrel{=}{
{\tiny (|\nu|<1)}}}
\sum_{n\geq 1}{\frac{(-1/6)_n(-5/6)_n}{n!^2}n\nu^n}~~.
\ee
Closing $\gamma_2$ toward $+\infty$ gives:
\be
\label{U_20}
U_2(\nu) ~_{\stackrel{=}{
{\tiny (|\nu|<1)}}}1+U_1(\nu)\ln \nu +\Phi(\nu)~~,
\ee
where:
\be
\Phi(\nu)=\sum_{n\geq 1}{\frac{(-1/6)_n(-5/6)_n}{n!^2}{n\nu ^n}
\left[\psi(n-1/6)+\psi(n-5/6)-2\psi(1)+\frac{1}{n}-2\sum_{j=1}^n{\frac{1}{j}}
\right]}~~.
\ee
Here $\psi(z):=\frac{d}{dz}\ln \Gamma(z)$ is the logarithmic derivative of the
$\Gamma$ function.

The expansions of $U_1$ and $U_2$ for $|\nu|>1$ can be obtained by closing the
contours $\gamma_j$ toward $-\infty$. This gives:
\bea
\label{U_inf}
U_1(\nu)&=&e^{\frac{i\pi}{6}}\Phi_1(\nu)-e^{-\frac{i\pi}{6}}\Phi_2(\nu)\\
U_2(\nu)&=&\frac{\pi}{\sin \frac{\pi}{6}}
\left[\Phi_1(\nu)+\Phi_2(\nu)\right]~~,\nn
\eea
where:
\bea
\Phi_1(\nu)&:=&
\frac{5}{24}\frac{\sqrt{3}}{\pi}
\frac{\Gamma(5/6)^2}{\Gamma(2/3)}\nu^{1/6} ~_{2}F_1
\big({\scriptsize \begin{array}{ccc}-1/6&~&5/6 \\~&5/3&~ \end{array}\big|\frac{1}{\nu}}\big)\nn\\
&=&\frac{1}{\Gamma(-1/6)\Gamma(-5/6)}\sum_{n\geq 0}{
\frac{\Gamma(n-1/6)\Gamma(-n-2/3)}{n! \Gamma(-n+1/6)}\nu^{-n+1/6}}~~\\
\Phi_2(\nu)&:=&-\frac{1}{6}
\frac{\Gamma(2/3)}{\Gamma(5/6)^2}\nu^{5/6} ~_{2}F_1
\big({\scriptsize\begin{array}{ccc}1/6&~&-5/6 \\~&1/3&~ \end{array}}\big|\frac{1}{\nu}\big)\nn\\
&=&\frac{1}{\Gamma(-1/6)\Gamma(-5/6)}\sum_{n\geq 0}{
\frac{\Gamma(n-5/6)\Gamma(-n+2/3)}{n! \Gamma(-n+5/6)}\nu^{-n+5/6}}~~.
\eea

To find the expansions for $|1-\nu|<1$, we first notice that the Picard-Fuchs
equation (\ref{PF}) admits the symmetry:
\be
\label{sym}
\nu\rightarrow 1-\nu~~.
\ee
Defining $U(\nu):=\left[\begin{array}{c}U_1(\nu)\\U_2(\nu)\end{array}\right]$,
we must therefore have:
\be
\label{S_relation}
U(1-\nu)=J U(\nu)~~,
\ee
for some constant involutive matrix $J$. Direct computation easily gives:
\be
J=\left[\begin{array}{cc} 0&\frac{1}{2\pi}\\2\pi&0\end{array}\right]~~,
\ee
which indeed satisfies $J^2=Id$. Together with (\ref{S_relation}) and
(\ref{U_10}), (\ref{U_20}), this specifies the expansions of $U_j$ for
$|1-\nu|<1$. The symmetry (\ref{sym}) interchanges the two 
singular points $\nu=0$ and $\nu=1$. This corresponds to the isomorphism 
(\ref{isomf}) between the two degenerations of the regularized curve 
(\ref{reg_curve}).

\subsection{Monodromies in the Meijer basis}

Let us define monodromy matrices around $\nu=0$ and $\nu=\infty$ by the 
relations:
\bea
U(e^{2\pi i}\nu)&=&T[0]U(\nu)~~{\rm for}~~|\nu|<<1\nn\\
U(e^{2\pi i}\nu)&=&T[\infty]U(\nu)~~{\rm for}~~|\nu|>>1~~.
\eea
With a similar definition of the monodromy matrix $T[1]$ around $\nu=1$, we
have:
\be
\label{hty}
T[\infty]=T[1]T[0]~~,
\ee
which results from a similar relation in the fundamental group of the moduli
space ${\cal M}=\mathbb{P}^1-\{0,1,\infty\}$. Using the results of the previous
subsection, one easily computes:
\be
T[0]=\left[\begin{array}{cc} 1&0\\2\pi i&1\end{array}\right]~~,~~
T[1]=\left[\begin{array}{cc} 1&\frac{i}{2\pi}\\0&1\end{array}\right]~~,~~
T[\infty]
=\left[\begin{array}{cc} 0&\frac{i}{2\pi}\\2\pi i&1\end{array}\right]~~,
\ee
where we used (\ref{hty}). Also notice the relation:
\be
T[1]=JT[0]J~~,
\ee
which holds as a consequence of (\ref{S_relation}).

\subsection{The integral basis}

Choosing the canonical basis $(A,B)$ as above, we define the periods:
\be
S=\frac{i}{2}\int_{A}{\frac{dx}{2\pi i}y}~~{\rm and } ~~
\Pi=\frac{i}{2}\int_{B}{\frac{dx}{2\pi i} y}~~.
\ee
It is easy to see that these are related to the Meijer periods
by the rescalings:
\bea
\label{SP}
S&=&\frac{36}{5\times 27}\frac{U_1}{\epsilon^2}\\
\Pi &=&\frac{1}{2\pi i} \frac{36}{5\times 27}\frac{U_2}{\epsilon^2}\nn
\eea
In the canonical basis, the monodromies take the form:
\be
\label{integral_mons}
{\tilde T}[0]=\left[\begin{array}{cc} 1&0\\1&1\end{array}\right]~~,~~ {\tilde
T}[1]=\left[\begin{array}{cc} 1&-1\\0&1\end{array}\right]~~,~~ {\tilde
T}[\infty] =\left[\begin{array}{cc} 0&-1\\1 &1\end{array}\right]~~,
\ee
while the matrix $J$ is replaced by:
\be
\label{tildeJ}
{\tilde J}=
\left[\begin{array}{cc} 0&i\\-i&0\end{array}\right]~~,
\ee
The first two monodromies have the Picard-Lefschetz form. 
In the limit $\nu=\infty$, both periods of the elliptic curve blow up.

\subsection{The flux superpotential}

Given the periods (\ref{SP}), one can now 
compute the flux superpotential of \cite{Cachazo}
\be
W_{eff}=-2\pi i \left[ N\Pi + \alpha S\right]~~
\ee
everywhere on the moduli space of the geometrically regularized model.
Of this moduli space, the regions 
of interest for the field theory applications are the vicinities of the 
two conifold points. We now show how the leading contributions to the 
flux superpotential around these points 
can be recovered from the periods (\ref{SP}), 
thereby yielding the Veneziano-Yankielowicz superpotential in our 
regularization.

\subsubsection{The flux superpotential for $|\nu|<<1$}

For small $\nu$, we have:
\bea
U_1&=&\frac{5}{36}\nu +O(\nu^2)\\
U_2&=&1+\frac{5}{36}\nu\left[\ln \nu -1-\ln(16\times 27)\right]
+O(\nu^2)~~,\nn
\eea
where we used the identity:
\be
\psi(1/6)+\psi(5/6)-2\psi(1)=-\log(16\times 27)~~.
\ee
This gives:
\bea
S&\approx &\frac{\mu}{4}\nn\\
\Pi& \approx&\frac{1}{2\pi i}\left[\frac{36}{5\times 27}\frac{1}{\epsilon^2}
+S(\ln\frac{\epsilon^2}{16}+\ln S -1)\right]~~.
\eea
Defining $\Lambda_0$ and $\Lambda$ through:
\bea
\Lambda_0^3&=& \frac{16}{\epsilon^2}\nn\\
\Lambda^{3N}&=&\Lambda_0^{3N}e^{-2\pi i \alpha}~~,
\eea
we easily obtain:
\be
W_{eff}=-\frac{N\Lambda_0^3}{60}+W_{VY}(S,\Lambda)+O(1/\Lambda)~~,
\ee
where:
\be
W_{VY}(S,\Lambda)=S\ln \frac{\Lambda^{3N}}{S^N} +NS 
\ee
is the Veneziano-Yankielowicz superpotential. 

\subsubsection{The flux superpotential for $|1-\nu|<<1$}

This results immediately from the above upon
using the symmetry (\ref{S_relation}). Defining:
\be
\label{exchange}
{\hat N}=-\alpha~~,~~{\hat \alpha}=+N~~,
\ee
we have:
\be
W_{eff}=\frac{{\hat N}{\hat \Lambda}_0^3}{60}+W_{VY}(\Pi,{\hat \Lambda})+
O(1/{\hat \Lambda})~~,
\ee
where:
\be
{\hat \Lambda}^{3{\hat N}}=
{\hat \Lambda}_0^{3{\hat N}}e^{-2\pi i {\hat \alpha}}~~,
\ee
with:
\be
{\hat \Lambda}_0^3=-\frac{16i}{\epsilon^2}~~.
\ee 
Relation (\ref{exchange}) corresponds to the $SL(2,\Z)$ transformation:
\be
\left[\begin{array}{c}{\hat \alpha}\\{\hat N}\end{array}\right]=
\left[\begin{array}{cc}0&1\\-1&0\end{array}
\right]\left[\begin{array}{c}\alpha\\N\end{array}\right]~~,
\ee
which accompanies the transformation (\ref{tildeJ}) 
on the periods $S,\Pi$. This agrees with the discussion in Subsection 3.1.

\section{Summary and outlook}

By considering a geometric regularization of the local conifold, we 
obtained a set of periods defined in terms of compact cycles. The
regularization preserves the $SL(2,\Z)$ invariance of the flux superpotential 
and allows one to determine this quantity at every point on the moduli space
by making use of the standard technique of Picard-Fuchs equations.

The regularized geometry (\ref{reg_conifold}) can be viewed as a monodromic 
$A_1$ fibration in the sense of \cite{Cachazo3}. 
This means that the resulting string background 
cannot be produced by a geometric transition 
from a background with wrapped $D5$-branes. 
Indeed, turning off the deformation parameter $\mu$ one finds local conifold
degeneration at $x=y=s=t=0$. One could blow up this point to a
two-sphere. However, there is a non-trivial monodromy around
the point $x_3=-\frac{1}{\epsilon}$ produced by the regulator $\epsilon$, 
and going around this point in the $x$-plane takes the 
two sphere to minus itself. As explained in \cite{Cachazo3}, D-branes can only
be wrapped on monodromy-invariant cycles of a fibered $ADE$-geometry.

However, the regularized geometry is 
a valid flux background of IIB string theory and as such it leads to a
non-trivial ${\cal N}=1$ superpotential in four dimensions. This flux 
superpotential is explicitly $SL(2,\Z)$ invariant.  An interesting feature 
of this construction is that the regularized space admits two conifold 
degenerations, which are achieved for different values of the deformation 
parameter $\mu$.  At the first conifold point (which occurs for $\mu=0$), 
the A-cycle vanishes. There one can perform a double scaling limit 
which recovers the usual geometry of \cite{Vafa_ts, Cachazo}. From
this point of view the effective four dimensional description which arises in
the limit $\alpha'\rightarrow 0$ of the IIB string theory 
serves as a sort of $SL(2,\Z)$ invariant 
completion of the strongly coupled ${\cal N}=1$ gauge theory.
The superpotential (\ref{W_flux}) is only the genus zero part in the
genus expansion.  In order to decouple higher genus contributions,
which correspond to gravitational corrections, one can take
$N\rightarrow \infty$ while keeping $N g_s$ fixed, 
where $g_s$ is the string coupling in the $SL(2,\Z)$ frame where the
$A$-cycle carries RR flux. 

As we have seen, the regularized geometry has a second conifold degeneration 
for $\mu=-\frac{4}{27\epsilon^2}$.  
There it is the B-cycle (the cycle carrying NS-NS flux) 
which shrinks to zero
size.   
By explicitly calculating the periods, we found that mathematically 
this second conifold point is completely equivalent to the first. Physically, 
this can be understood by performing an $SL(2,\Z)$ duality
transformation of the IIB string, which exchanges the RR and NS-NS 
sectors, while inverting the string coupling 
$g_s\rightarrow g'_s = \frac{1}{g_s}$.  One can then
go trough the same steps as before, by defining ${\hat N}=-\alpha$ and 
${\hat \alpha}=N$ and performing a double scaling limit which
keeps the B-period finite while taking 
${\hat N}\rightarrow \infty$ with ${\hat N} g'_s$ fixed. 
From the point of view of the original $SL(2,\Z)$ frame, this is a strong
coupling limit. That leading terms in 
the effective superpotential still take the
Veneziano-Yankielowicz form in this limit is not surprising, since as a
holomorphic quantity it is protected and thus can be computed at strong
string coupling with the help of NS-NS-flux instead of RR-flux.  Hence the
geometry (\ref{reg_conifold}) provides us with a manifestly $SL(2,\Z)$
invariant flux background. 

It is straightforward to extend the geometric regularization to
more complicated cases, e.g. for multi cut situations corresponding to a
breaking of the gauge group to $n$ factor subgroups, 
which are engineered by the 
space (\ref{cy}). In this case, the regularization replaces (\ref{cy}) with: 
\be
\label{nreg}
\epsilon x^{2n+1} + W'(x)^2 + y^2 + t^2 + s^2 =0~~.
\ee
Reducing this in the manner of 
\cite{Klemm:1996bj, Cachazo, Cachazo2, Cachazo3} 
gives a genus $n$ hyperelliptic Riemann
surface, thus introducing a new branch point 
and a new cut connecting it with the point at infinity.  
Again the reduction
$\omega$ of the holomorphic three-from of (\ref{nreg}) produces a meromorphic
differential of the second kind. 
There will be an $Sp(2n,\Z)$ symmetry acting on
a symplectic basis of $A$ and $B$ cycles which together with type
IIB $SL(2,\Z)$ duality transformations should give rise to various
dual superpotentials. Another possibility is that two or more cycles with
non-vanishing intersection form vanish at a point at finite distance in the
moduli space. It would be interesting to study if this happens and what the
physical interpretation could be.  Similar methods could also be applied for
generalizations based on ADE fibrations \cite{Cachazo2, Cachazo3,Radu2} or
orientifolds \cite{Radu}. Finally, let us mention that the geometric
regularization could prove useful in the study of flux
backgrounds with orientifolds \cite{orientif}, where one also encounters
`noncompact cycles' when using a cutoff regularization. Since the geometric 
regularization allows one to apply standard Picard-Fuchs techniques, it 
could also be useful for explicit computations of periods in a large number of 
situations involving non-compact Calabi-Yau spaces.

{\bf{Acknowledgment}}: We would like to thank A. Klemm for useful discussions.
This work has bee supported by DFG-project KL1070/2-1.

\end{document}